\documentclass{basi}
\usepackage{graphicx}
\def\lsim{\lower.5ex\hbox{$\; \buildrel < \over \sim \;$}}
\def\gsim{\lower.5ex\hbox{$\; \buildrel > \over \sim \;$}}
\def \simeq{\lower.3ex\hbox{$\; \buildrel \sim \over - \;$}}
\def\ch{\lower-0.55ex\hbox{--}\kern-0.55em{\lower0.15ex\hbox{$h$}}}
\def\lh{\lower-0.55ex\hbox{--}\kern-0.55em{\lower0.15ex\hbox{$\lambda$}}}

.tfm

\begin{document}
\title[Numerical Simulation of Recombination of Hydrogen]{Recombination Efficiency of Molecular Hydrogen 
on Interstellar Grains-II A Numerical Study}

\author[Sandip K. Chakrabarti, Ankan Das, Kinsuk Acharyya and Sonali Chakrabarti]
{Sandip K.\ Chakrabarti$^{1,2}$,  Ankan Das$^{2}$, Kinsuk Acharyya$^2$ and Sonali Chakrabarti$^{3,2}$\\
$^1$S.N. Bose National Center for Basic Sciences, JD-Block, Salt Lake, Kolkata, 700098; e-mail: chakraba@bose.res.in \\
$^2$ Centre for Space Physics, Chalantika 43, Garia Station Rd., Kolkata, 700084; e-mail: ankan@csp.res.in\\
$^3$ Maharaja Manindra Chandra College, 20 Ramkanta Bose Street, Kolkata 700003; e-mail: sonali@csp.res.in\\ }

\maketitle

\begin{abstract}
A knowledge of the recombination time on the grain surfaces has 
been a major obstacle in deciding the production rate of molecular
hydrogen and other molecules in the interstellar 
medium. We present a numerical study to compute 
this time for molecular hydrogen for various cloud and grain 
parameters. We also find the time dependence, particularly when 
a grain is freshly injected into the system.
Apart from the fact that the recombination 
times seem to be functions of the grain parameters such as 
the activation barrier energy, temperature etc, our result also shows the dependence  
on the number of sites in the grain $S$ and the effective accretion rate per site $a_s$ of atomic 
hydrogen. Simply put, the average time that a pair of atomic hydrogens
will take to produce one molecular hydrogen depends on how heavily the grain is
already populated by atomic and molecular hydrogens and how fast the
hopping and desorption times are. We show that if we write the average
recombination time as $T_r \sim S^\alpha/A_H$, where, $A_H$ is the hopping rate,
then $\alpha$ could be much greater than $1$ for all astrophysically relevant accretion 
rates. Thus the average formation rate of $H_2$ is also dependent on the
grain parameters, temperature and the accretion rate.
We believe that our result will affect the overall rate of the formation 
of complex molecules such as methanol which require successive hydrogenation on the
grain surfaces in the interstellar medium.
\end{abstract}


\section{Introduction}

It has long been suggested that the dust grains play a major role
in the formation of molecular hydrogen in the interstellar medium
(ISM) (Gould \& Salpeter, 1963). Considerable studies were made since
then to understand the real physical processes which are taking
place both theoretically (e.g., Hollenbach, Werner \& Salpeter, 1971;
Takahashi, Matsuda \& Nagaoka, 1999; Biham et al. 2001) as well as
experimentally (e.g., Pirronello et al. 1997a,b, 1999). More recently,
Biham et al. (2001), and Green et al. (2001) have computed $H_2$
production rate by physisorption. It was found that a significant production
is possible in cooler ($\sim 10-25$K) clouds. Cazaux \& Tielens 
(2002, 2004) use both physisorption and chemisorption, to demonstrate
that $H_2$ production is possible at high temperatures ($\sim 200-400$K) also.
The goal is to study the rate at which the $H$ atoms combine together
on the surface of the grains to form $H_2$ and then they are desorbed
into the gas phase to react with other atoms. When compared with the average
mass fractions of various molecular species obtained through gas phase reactions
(see, Chakrabarti \& Chakrabarti, 2000ab; Das et al. 2006), it was found that the observed abundances
of more complex species, such as methanol, are much higher.
It is possible that methanol as well as its precursors also have to be formed on grain surfaces
through successive hydrogenation. Our finding for molecular hydrogen has thus important 
bearings on the formation of more complex molecules on grains. These molecules would, in turn,
desorb into the gas phase and would be expected to produce more complex species 
such as amino acids in due course.

One of the most challenging problems is to determine the average rate
at which the recombination of atomic hydrogen takes place on a grain surface.
In theoretical investigations which are prevalent in the subject (See, Acharyya and 
Chakrabarti, 2005; hereafter Paper I;  Acharyya, Chakrabarti and Chakrabarti, 2005; hereafter ACC05),
the diffusion rate $A_H$ (inverse of the diffusion time $T_d =1/A_H$)
is divided  by $S$, the number of sites on the grain surface 
(e.g., Biham et al. 2001) to get the recombination rate.
The argument for reducing the rate by a factor of $S$ is this: on an
average, there are $S^{1/2}$ number of sites in each direction of the
grain. Since the hopping is random, it would take square of this,
i.e., $S$ number of hopping to reach a distance located at $S^{1/2}$ sites away, where,
on an average, another $H$ is available. Thus, the effective recombination
rate was chosen to be $A_H/S$. It is an empirical factor and needs
more careful treatment. In our present paper, we replace $S$ by an 'unknown' 
quantity $S^\prime = S^\alpha(t)$, where $\alpha(t)$ may be time dependent
(if the grain and cloud parameters change) and it could also deviate from unity.
Let $\alpha=\alpha_0$ when a steady state is reached. Higher the accretion rate,
lesser should be the value of $\alpha_0$ as the effective surface area   
$S^\prime$ $(t \rightarrow \infty)$ gets smaller and smaller. 
The opposite is true for smaller accretion rates. In fact, in the limit, if the
accretion rate is so low, that a lone $H$ sweeps around the grain several times to
find another $H$, one would get $\alpha_0>1$ since the effective site number
is higher than $S$. Using our simulation, we 
determine how the effective site number deviates from $S$, one way or the other
when the accretion rate is varied. Our result is likely to have important consequences for the 
formation of other hydrogenated species, such as water, methanol on grain surfaces. 
This will be discussed elsewhere. Some preliminary results with steady state $\alpha_0$
have been presented in Das et al. (2005) 
and  Chakrabarti et al. (2006). In the current paper, we discuss the time
and temperature dependence of $\alpha (t)$ and studied the cases for more varied astrophysically important
accretion rates.

In the next Section, we present the modified typical equations which govern the 
molecular hydrogen production rate on a grain surface. Incorporating the 
{\it physical aspects} of these equations, we perform 
a numerical simulation to determine the numbers on the grains. In \S 2, 
we present the procedure for the simulation and in \S 3 we present the 
results for two types of commonly used grains, namely, olivine and amorphous 
carbon. We show how $\alpha(t)$ depends on time and how it settles 
into a number (generally, $>1$) when steady state is reached. Finally, in \S 4, 
we present our concluding remarks.

\section{Procedure of numerical simulation}

The relevant equations which are generally solved on grain surfaces have been
presented in Paper I (Eqs. 2a-d) and we do not repeat them here. However,
for Monte-Carlo simulation we need to modify these equations.
Since in our simulation we expect to get the rate of diffusion 
($T_r(t)=S/A_H$) of H exactly, we  can {\it assume} that the recombination
time at any instant $t$ could be written in the form:
$$
T_r(t) = S^{\prime} /A_H ,
\eqno{(1)}
$$
where, $S^\prime$ is an `effective surface area' which we may be written as 
$S^\prime=S^\alpha(t)$.
Here, $\phi_H=F_H (1-f_{grh}-f_{grh2})$.  $F_H$ is the accretion rate of $H$ and
$F_H= A <v> N_H$, $A$ being the area of a grain, $<v>$ is the average velocity
and $N_H$ is the number density of $H$ in the gas phase. Our goal would 
be to compute the steady state value $\alpha_0 =\alpha(t \rightarrow \infty)$
for various grains (type and size) at various accretion rates and grain 
temperatures and to check if $\alpha \sim 1$. Thus, instead of Eqs. 2a of Paper I,
the actual governing equation for atomic hydrogen should be,
$$
\frac {dn_H}{dt}=\phi_H - W_H n_H -2 (A_H/S^\alpha)  n_H^2,
\eqno{(2a)}
$$
along with equation governing the rate of production of $H_2$ given by,
$$
\frac {dn_{H2}}{dt}=\mu (A_H/S^\alpha) n_H^2 - W_{H2} n_{H2},
\eqno{(2b)}
$$
where, $A_H$ is the hopping rate given by $\nu\exp(-E_0/k_bT)$,  
$W_H$ is the desorption co-efficient of hydrogen atom given by,
$\nu\exp(-E_1/k_bT)$, $W_{H2}$ is the desorption co-efficient of hydrogen 
molecule given by, $\nu\exp(-E_2/k_bT)$. Here, 
$E_0$ is the activation barrier energy for diffusion of H atom,
$E_1$ is the activation barrier energy for
desorption of $H$ atom and $E_2$ is the activation barrier energy for
desorption of $H_2$ molecule, $\nu$ is the vibrational frequency given by,
$\nu= \frac{2 s E_d}{\pi^2 m_H}$,  $n_H$ be the number of $H$ 
atoms, and $n_{H2}$ be the number of $H_2$ on a grain at time $t$,
$s \sim 10^{14-16}$ is the surface density of sites on a grain,
$m_H$ is the mass of the $H$ atom and $E_d$ is the binding energy. 
The parameter $\mu$ represents the fraction
of $H_2$ molecules that remains on the surface upon formation while (1-$\mu$)
fraction is desorbed due to the energy released in the recombination
process. The $H_2$ production rate $R_{H2}$ in the gas due to grain is then given by,
$$
R_{H2}=(1-\mu) (A_H/S^\alpha) n_H^2 + W_{H2}  n_{H2}.
\eqno{(3)}
$$
The values of the activation barrier energies $E_0$, $E_1$ and $E_2$ are
taken from Katz et al. (1999) and are given by, $E_0 = 24.7$ meV,
$E_1 = 32.1$ meV and $E_2 = 27.1$ meV for olivine
and $E_0 = 44$ meV, $E_1 = 56.7$ meV and $E_2 = 46.7$ meV
for amorphous carbon grains. For olivine,  $\mu = 0.33$
and for amorphous carbon $\mu = 0.413$ was used.

For the sake of simplicity, we assume each grain to be square in shape having
$S=n^2$ number of sites (square lattice). We use periodic boundary condition so that
the lattice behaves like a closed surface. We anticipate that the result would be
insensitive to the actual nature of the lattice when simulation is done for a long time.
We use mono-layer on the grain surface. We choose the minimum time step to
be $t_h=1/A_H$, the hopping time and advance the global time by this step. 
If $\phi_H$ is the effective accretion rate (as defined
above), after every time step, i.e., after every $1/A_H$ seconds,
$\phi_H/A_H$ number of hydrogen atoms are dropped on the grain. If
$\phi_H$ is too low so that $\phi_H/A_H <1$, then, clearly, one H is dropped after 
every $A_H/\phi_H$ steps. The exact site at which one atom is dropped is
obtained by a pair of random numbers ($R_x, R_y$, $R_x, R_y <1$) obtained by a random number
generator. This pair would place the incoming hydrogen at ($i, j$)th grid,
where, $i$ and $j$ are the nearest integers obtained using the Int function: $i=int(R_x*n+0.5)$
and $j=int(R_y*n+0.5)$. Each atom starts hopping
with equal probability in all four directions. This was decided by
another random number. When, during the hopping process, one atom enters a site
which is already occupied by another atom, we assume that a molecule has
been formed and increase the number of $H_2$ by unity. However, when the
atom enters a site occupied by an $H_2$, another random number is generated
to decide which one of the other neatest sites it is going to occupy.

Thermal evaporation of $H$ and $H_2$ from a grain surface are handled in the
following way. Since $W_H$ is the desorption rate for $H$, one atom is supposed to
be released to the gas phase after every $1/W_H$ seconds. We generate a random number $R_t$ for every
$H$ present on the grain and release (at each time step, i.e., $1/A_H$ seconds)
only those for which $R_t< W_H/A_H$. Similar procedure is followed for the evaporation
of $H_2$ for which the criterion for evaporation was $R_t < W_{H2}/A_H$.

Due to spontaneous desorption, a factor of (1-$\mu$) of $n_{H2}$ is
lost to the gas phase. Here too, a random number $R_s$ is
generated for each newly formed (within that time step) $H_2$
present on the grain. Those which satisfy $R_s < (1-\mu)$ are removed to the gas phase.

We continue our simulation for more than $10^9$s. After an initial transient
period, $n_H$ and $n_{H2}$ reaches a quasi-steady state with some
fluctuations. In a steady state, $\alpha_0$ can be calculated using Eq. 2a, i.e.,
$$
\alpha_0= log(\frac{2A_H n_H^2}{\phi_H-W_Hn_H})/log(S) .
\eqno{(4)}
$$
In this context, one important parameter $\beta$ may be defined as
the `catalytic capacity' of a grain (Chakrabarti et al. 2006) which measures the efficiency of the
{\it formation} of $H_2$ on that grain surface for a given pair of $H$ residing on it.
Let $\delta N_{H_2}$ be the number of $H_2$ {\it formed} in $\delta t$ time.
Since two hydrogen atoms are required to create one $H_2$,
the average rate of creation of one $H_2$ per pair of $H$ atom would be given by,
$$
<A_{H1}> = \frac{1}{2n_H} \frac{\delta N_{H_2}}{\delta t}.
\eqno{(5a)}
$$
We identify the inverse of this rate with the average formation rate given by,
$$
T_f(t) = S^{\beta(t)} /A_H .
\eqno{(5b)}
$$
Thus,
$$
S^\beta(t) = A_H/<A_{H1}>.
\eqno{(5c)}
$$
This yields $\beta(t)$ as a function of time as,
$$
\beta = log(A_H/<A_{H1}>)/log(S).
\eqno{(6)}
$$
Here too we can define $\beta_0=\beta(t \rightarrow \infty)$.
Our goal would be to see if $\beta_0$ thus obtained actually varies with grain parameters, accretion rate
and temperatures. Intuitively, it should. This is because, when the accretion rate is very low,
finding a second $H$ on the grain would be difficult and it can take several sweeps of the grain surface.
Thus $\beta_0 >1$ would be a possibility. On the other hand, when the rate is very high, many sites
would be occupied and another $H$ would be met along any direction and $\beta_0$ could come down to
$1/2$ or even less. Indeed, this is what we see as well. We have carried
out the simulation with and without spontaneous desorption taken into account. In presence of the 
spontaneous desorption, we remove $H_2$ by generating a random number as soon as
one $H_2$ is formed and checking if it is less or more compared to $\mu$. If less, $H_2$ 
remains on the grain, else it is taken out to the gas phase.
The detailed results are discussed in the next Section.

\section{Results and Discussions}

Fig. 1 gives the variation of $n_H$ and $n_{H2}$ residing on the grain surface
since the beginning of the simulation. We chose a grain with $10^4$ sites at $8$K. 
For clarity, we plot average numbers in every $\sim 5.5 \times 10^6$s bin after the 
initial transient period of $\sim 2 \times 10^6$s is over.
The simulation was carried out near about  $10^{10}$s. The effective accretion rate is
assumed to be $\phi_H=7.98 \times 10^{-4}$s$^{-1}$.
In this case, the steady state has clearly been reached by this time and
we can compute $\alpha_0$ by using Eq. 4. This is done below.

\begin {figure}
\vskip 0.5cm
\includegraphics[width=5.1cm]{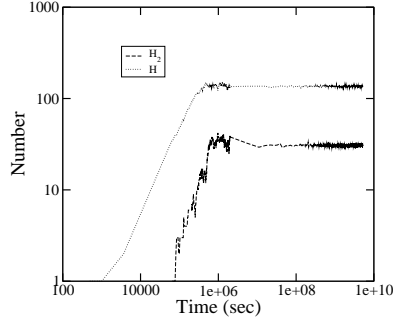}
\centering
\vskip 0.5cm
\caption{: Variation of the number of H and $H_2$ on an olivine grain of
$10^4$ sites kept at $8$K and exposed to an effective accretion rate per site
of H $7.98 \times  10^{-8}$ $s^{-1}$. We carried out our simulation near 
about $10^{10}$s. For cleanliness, after the initial transient period of 
about $2 \times 10^6$s is over, we took time average at every $5.5\times 10^6$s before plotting
the numbers. }
\end {figure}

\subsection {Olivine grains}
We start with the olivine grains which are kept at 8 K.
The evaporation rate of $H_2$ is $W_{H_2}= 0.834844 \times 10^{-5}$ s$^{-1}$
which corresponds to a time scale of $\sim 119782.8$s and similarly the 
evaporation rate of $H$ is $W_H=0.59 \times 10^{-8}$s$^{-1}$ which corresponds to
 $\sim 169585908$s. The hopping time for
hydrogen on this grain is $3680.58$s (data taken from Katz et al. 1999).
                                                                                                                       
In Fig. 2a, we present the computed $\alpha_0$ (Eq. 4) as a function
of $a_s$ -- the effective accretion rate per site. The solid, dot-dashed
and the dashed curves are for $S=10^4$, $9\times 10^4$ and for
$10^6$ sites respectively. No spontaneous desorption has been
included (i.e., $\mu=1$). We extrapolated our curve to very low accretion rates
which would have taken a very long computation time, just to show the
trend of the result and extreme conditions. We note that $\alpha_0$ is generally higher
than unity in the region of our interest. This is because $H$s are scattered few 
and far between and it takes a longer time (generally more than 
one sweeping) for one $H$ to locate another. $\alpha_0$
monotonically drops as the accretion rate goes up. In Fig. 2b, we show the
variation of $\beta_0$ (Eq. 6) for the same case. Here too, we see that $\beta_0$
is very high compared to unity for low rates, but becomes $\sim 0.5$ or lower
for higher rates as expected. Note that $\alpha_0$ and $\beta_0$
go down with increasing site number $S$ also. Since for a smaller
grain, the possibility of getting it filled at a high rate is
higher, one would have expected an opposite result.  However, it is to be
remembered that for a larger grain, the accretion rate itself ($\phi_H=S a_s$)
is also large. Hence the plots are to be compared carefully.
For instance, the result of $a_s=0.0005\times 10^{-8}$$s^{-1}$ for
 $9\times 10^4$ sites is to be compared with that of $a_s=0.0045\times
 10^{-8}$s$^{-1}$ for $10^4$ sites in order to make a meaningful comparison.
In any case, for reasonable $\phi_H$ values with number densities up to $10^6$
cm$^{-3}$, the relevant $a_s$ would be below $10^{-6}$ where $\alpha_0>1$ in general.
                                                                                                                       
\begin {figure}
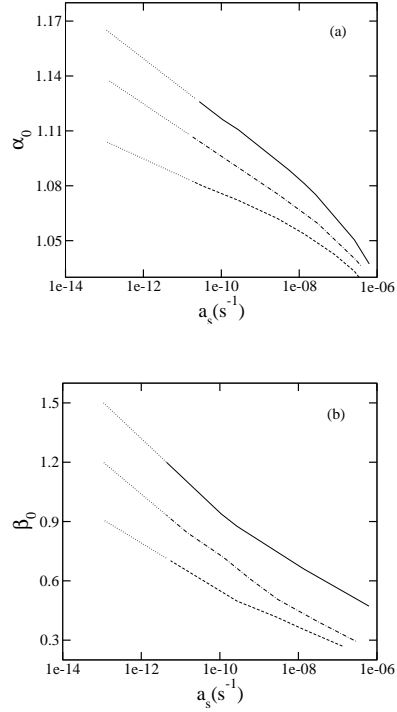

\vskip 0.5cm
\includegraphics[width=5.1cm]{fig2a.eps}
\centering
\vskip 0.7cm
\includegraphics[width=5.1cm]{fig2b.eps}
\vskip 0.5cm
\caption{\small{\bf (a-b)}: Variation of (a) $\alpha_0$ and (b) $\beta_0$ as a 
function of $a_s$, the effective accretion rate per site for various olivine
grains kept at $8$K. The dashed, dot-dashed
and solid curves are for $S=10^6,\ 9\times 10^4, and \ 10^4$ respectively.
 $\alpha_0$ is clearly a function of the 
accretion rate. For rates relevant in molecular clouds $\alpha_0$ 
and $\beta_0$ are much larger than unity. For very high rates $\beta_0$ 
comes down to $0.5$ or lower. The deviation is highlighted using 
dotted curves by extrapolating at very low accretion rates. }
\end{figure}

\begin {figure}
\includegraphics[width=12.7cm]{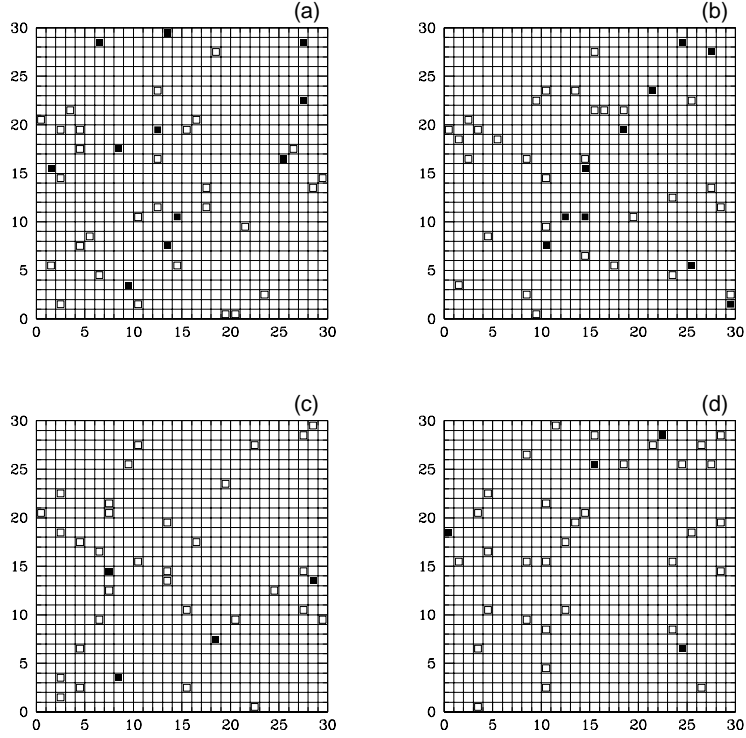}
\centering
\vskip -.9cm
\caption{\small{\bf (a-d)}: In (a-b) snapshots of the grain surface 
with $H$ (hollow squares) and $H_2$ (filled squares) at two arbitrarily 
chosen times (a) $ 8 \times 10^8$s and at (b)$ 10^9$s. Here an olivine 
grain (at $8$K) with $900$ sites has been chosen. This is bombarded with 
an accretion rate per site of H $3.02 \times 10^{-7}$ per sec. 
No spontaneous desorption has been assumed here. In (c-d) spontaneous 
desorption has been included and plotted for the same time 
as before. Thus, numbers of $H_2$ residing on the grain at 
any instant are lesser.}

\end{figure}


\begin{figure}
\vskip 0.5cm
\includegraphics[width=5.1cm]{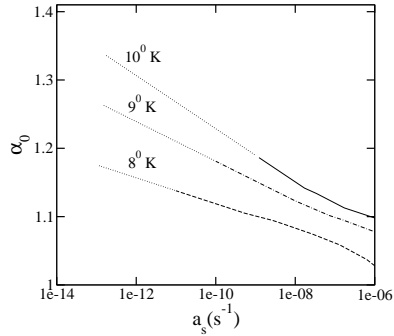}
\centering
\vskip 0.5cm
\caption{: Temperature dependence of $\alpha_0$ for the olivine grains at 
$10$K (solid), $9$K (dot-dashed) and $8$K (dashed). The deviation is highlighted 
using dotted curves by extrapolating at very low accretion rates.}
\end{figure}

\begin{figure}
\vskip 0.5cm

\includegraphics[width=5.1cm]{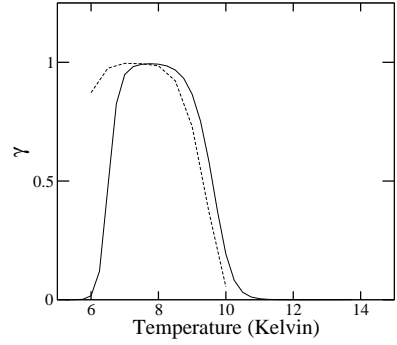}
\centering
\vskip 0.5cm
\caption{: A comparison between the recombination efficiency obtained from the 
rate equation (solid) and that obtained from our simulation 
(dashed). We use the accretion rate per site $1.8 \times 10^{-9}$ per second 
for a grain of diameter $0.1 \mu$m. 
The difference can be attributed to the temperature dependence of the 
$\alpha_0$ as shown in Fig. 4 above.}
\end{figure}

\begin{figure}
\vskip 0.5cm
\includegraphics[width=5.1cm]{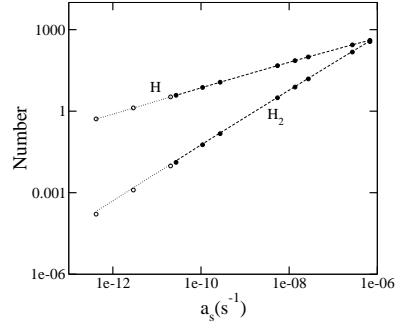}
\centering
\vskip 0.5cm
\caption{: A comparison of the simulation results (dark circles) with 
those obtained from analytical considerations (dashed curves)
when suitable modification of the average recombination rate is made.
An Olivine grain of $10^4$ sites at a temperature of $8$K has been chosen in 
this comparison. Dotted curves are drawn using analytical results for 
$\alpha_0$ extrapolated to very low accretion rates.}
\end{figure}
                                                                                                                       
In Fig. 3(a-b), we present snapshots of the occupancy of $H$ (hollow squares)
and $H_2$ (filled squares) at two instants of time on the grain containing only
$900$ sites at two arbitrarily chosen times (a) t=$8 \times 10^8$s (b) t=$10^9$s
respectively. Here an olivine grain at ($8$K) with $900$ sites has been chosen.
This grain is facing an accretion rate of $F_H/S$ of $3.02 \times 10^{-7}$s$^{-1}$ 
per site. No spontaneous desorption has been assumed here. This is to be 
compared with Fig. 3(c-d) for the same simulation and for the same time as before
 when the spontaneous desorption has been included. The numbers of $H_2$ 
are fewer since after formation of $H_2$ molecules some part of the $H_2$ 
are spontaneously desorbed in to the gas phase.
 
When temperatures of the grain is increased, all the rates go down exponentially. 
As a result, we expect $\alpha_0$ to rise with temperature for a given accretion rate.    
In Fig. 4 we show this behaviour for $T=8, \ 9, \ 10$K respectively for olivine grains. 
We choose $S=10^4$ in this case. This behaviour affects the recombination
efficiency $\eta$ as defined by,
$$
\eta=\frac{2R_{H_2}}{F_H}
\eqno{(7)}
$$
itself. In Fig. 5, we compare the temperature dependence of 
$\eta$ as obtained from the rate equations (solid) curve with that obtained
from our simulations. The accretion rate per site 
of $1.8 \times 10^{-9}$ sec$^{-1}$ was used (same in both the cases). 
We note that for  $T\lsim7.5$K  the simulation results are higher and 
for $T\gsim 7.5$K the simulation results are lower. This is because 
$\alpha_0$ itself is strongly temperature dependent as shown in Fig. 4 while
the rate equation uses $S=S^\prime$  (i.e., $\alpha_0=1$) for all temperatures.

In Fig. 6 we compare our results with the analytically obtained results from the 
rate equation method provided correct $S^\prime$ was chosen (Eqs. 2ab). 
The simulation results are shown by the dark circles and 
those obtained from the analytical considerations are shown by the dashed curves.
An Olivine grain of $10^4$ sites at a temperature of $8$K has been chosen in 
this comparison. Dotted curves are drawn using analytical results for 
$\alpha_0$ extrapolated to very low accretion rates.

It is interesting to compare the results of our simulation with those
obtained from the analytical considerations with and without our 
$\alpha_0$ factor. In Table 1, we present this comparison. We take an Olivine grain of $10^4$ sites at $8$K and vary 
the accretion rates. In Column 1, we give the accretion rate per site of the grain. In Column 
2 we present the coefficient $\alpha_0$ which we derive from our simulation.  
In Columns 3-5, we present the number of $H$ as obtained by our simulation and the
modified equation (Eq. 2a) and the standard equations (Paper I) respectively. Columns 6-8,
we present similar results for $H_2$. We find that our simulation matches more accurately with 
the analytical results provided $S^\prime$ is chosen as the surface area. 
If the standard equation is used, the deviation is very significant. Indeed,
the number of $H$ on the grain could be roughly half as much when 
simplistic analytical model is used. What observe is that on the grains
we tend to have more $H$ and less $H_2$ than what analytical work suggests.

\begin{table}
\caption{\label{table} Comparison of $H$ and $H_2$ abundances in various methods}
\vskip 0.5cm
\hskip -1.2cm
\begin{tabular}{|ll|lll|lll|}
\hline
 Accretion Rate&  $\alpha_0$ &  & $H$ with &  &   & $H_2$ with  &  \\ \cline{3-5} \cline{6-8}
 per site  $A_s(S^{-1})$ & & simulation &  $\alpha_0\ne 1$ & $\alpha_0=1$ & simulation & $\alpha_0\ne 1$& $\alpha_0=1$\\
\hline\hline

$  6.79 \times 10^{-7}$ &  1.04 & 403.11&  407.80&  340.57&  361.71&  383.82 & 377.48\\
$  2.72  \times 10^{-7}$&  1.05&  275.78&  278.02&  219.31 & 149.96&  158.16&  156.53\\
$ 2.72 \times 10^{-8}$ & 1.07 & 99.03 & 99.31 & 70.35 & 15.40 & 16.14 & 16.11\\
$  1.36 \times 10^{-8}$ & 1.08 & 72.27 & 72.42&  49.80&  7.71 & 8.08 & 8.07\\ 
$  5.43 \times 10^{-9}$& 1.09 & 47.30&  47.37 & 31.52&  3.11 & 3.23 & 3.23\\
$  2.72 \times 10^{-10}$&  1.11 & 11.62&  11.62&  7.01&  $1.49 \times 10^{-1}$ & $1.59 \times 10^{-1}$ 
& $1.60 \times 10^{-1}$\\
$  1.09\times 10^{-10}$ & 1.12 & 7.46&  7.48&  4.42 & $5.81 \times 10^{-2}$ &  $6.24 \times 10^{-2}$
& $ 6.35\times 10^{-2}$\\
$  2.72 \times 10^{-11}$&  1.13 & 3.86 & 3.82 & 2.18 & $1.31 \times 10^{-2}$ & $1.49 \times 10^{-2}$ & $1.55 \times 10^{-2}$\\
$^*  2.04\times 10^{-11}$ & 1.13 & 3.36&  3.33 & 1.89 & $9.7 \times 10^{-3}$ & $1.11 \times 10^{-2}$ & $1.17 \times 10^{-2}$\\
$^*  2.92\times 10^{-12}$ & 1.14 & 1.32&  1.30 & 0.70 &  $1.3 \times 10^{-3}$&  $1.5 \times 10^{-3}$ & $1.6 \times 10^{-3}$\\
$^*  4.16\times 10^{-13}$ & 1.16& 0.52 & 0.51&  0.26 &  $2.00 \times 10^{-4}$&  $2.00 \times 10^{-4}$&  $2.00 \times 10^{-4}$\\
$^*  5.95\times 10^{-14}$ & 1.17 & 0.20 & 0.20&  0.10  & $2.11 \times 10^{-5}$&$  2.81\times 10^{-5}$& $ 3.13 \times 10^{-5}$\\
\hline\hline
\end{tabular}
\vskip 0.3cm
$*$ represents the extrapolated value.
\end{table}

\subsection {Amorphous carbon grains}

Carbon grains produce significant $H_2$ at a higher temperature than olivine
grains because all the barrier energies are higher. We plot in Fig. 7a the
variation of $\alpha_0$ with $a_s$ at temperature $14$K. The nature of variation 
of $\alpha_0$ remains the same, namely, $\alpha_0$ goes down with $a_s$. The 
solid, dot-dashed and the dashed curves are for  $S=10^6$, $9\times 10^4$ and 
for $10^4$ sites respectively. In Fig. 7b, we show the variation of $\beta_0$ and 
as expected its value can become as low as $0.5$ for very large accretion rate. These
results are representative as they are strongly temperature dependent as in the
case of olivine (see, Fig. 4).

\begin{figure}
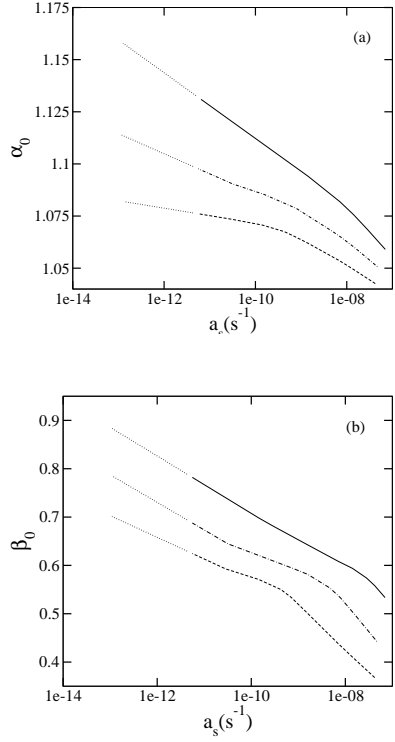

\vskip 0.5cm
\includegraphics[width=5.1cm]{fig7a.eps}
\centering
\vskip 0.7cm
\includegraphics[width=5.1cm]{fig7b.eps}
\centering
\vskip 0.5cm
\caption{\small{\bf(a-b)}: Variation of (a) $\alpha_0$ and (b) $\beta_0$ as a 
function of $a_s$, the effective accretion rate per site, for various 
amorphous carbon grains kept at $14$K. The solid, dot-dashed and
dashed curves are for $S=10^4, \ 9\times 10^4, \ 10^6$ respectively
The exponents $\alpha_0$ and $\beta_0$ are strong functions of the accretion 
rate. For low rates $\beta_0$ is higher than unity, while for higher rates 
it is close to $0.5$ or even lower. The deviation is highlighted using dotted
curves by extrapolating at very low accretion rates.} 
\end{figure}

\section {Concluding remarks}

In this paper, we studied the behaviour of the average recombination time of atomic hydrogens 
on grain surfaces as a function of time, temperature, accretion rate and grain parameters. 
In the literature, it is simply assumed that the recombination time is grain 
site number divided by the diffusion rate. However, we find that this 
simplistic assumption is not valid especially when the accretion 
rate of $H$ on the grain is very low or very high. For very low rates, an 
atomic hydrogen may have to sweep a grain several times before meeting
another atom to form a molecule. For very high rates, the grain surfaces could be
partially filled with $H$ and $H_2$ and search for another $H$ need not take $S$ number of
hops. We found that on an average, $\alpha_0$, the index which determines the average
recombination time, could be grater than 1 for very low rates. We also find that 
$\alpha_0$ depends very strongly on the nature of the grains as well as the temperature and grain site numbers. 
We show how the number of H and $H_2$ change with time and determine the time scales in which 
quasi-equilibrium is reached. Thus the recombination time is a complex function of these parameters. 
We defined another index $\beta$ (Chakrabarti et al. 2006) which is a measure of the average formation rate 
of $H_2$. This index also is also a strong function of accretion rate per site. In fact, for 
very high rates the index may go down to $0.5$ or even lower. A comparison of the recombination
efficiency as obtained from our procedure with that obtained from the rate equation clearly shows
a deviation easily attributable to the temperature dependence of $\alpha_0$ in a realistic case.

The results we obtained are likely to be astrophysically important because of several factors. The 
physical reason of the dependence of the average recombination time is so generic that one needs 
to revise all the abundances made on the grain surfaces. We find that  even for $H$,
we have more $H$ left over on grains, while less $H_2$ is produced and released. Thus,
analytical considerations will always over produce molecules obtained on grain surfaces.
Similarly, if, for example, methanols are produced through successive hydrogenation process on a grain surface, 
at each step such corrections would be required and the final abundance would be greatly affected. This
will be done in near future.
                                                                                                                       
Work of AD has been partly supported by a DST Project and the work of KA is partly 
supported by an ISRO project. SKC and AD also acknowledge Abdus Salam ICTP
where a part of the work has been completed while they visited as Senior Associate and
a conference participant respectively.

{}
                                                                                                                       
\end{document}